\documentclass[aps,prl,amsmath,amssymb,showpacs,floatfix,twocolumn]{revtex4}
\usepackage{graphicx}
\usepackage{comment}

\newcommand{\beq}{\begin{equation}}
\newcommand{\eeq}{\end{equation}}
\newcommand{\be}{\begin{eqnarray}}
\newcommand{\ee}{\end{eqnarray}}

\begin{document}

\title{Analytic Partition Function Zeros of the Wako-Saito-Mu\~noz-Eaton\\ $\beta$-hairpin Model}
\author{Julian Lee}
\email{jul@ssu.ac.kr}
\affiliation{Department of Bioinformatics and Life Science, Soongsil University, Seoul, Korea}
\date{\today}

\begin{abstract}
An {\it analytic} formula for the density of states of Wako-Saito-Mu\~noz-Eaton model, for simple classes of $\beta$-hairpins, is obtained. Under certain simplifying assumptions on the structure of the turn region and the values of local entropy, the partition function zeros are also obtained in {\it analytic} forms. The zeros are uniformly distributed on a circle, exhibiting  a first-order-like nature of the folding transition. After introducing a hydrophobic core at the central region of the hairpin, the zeros are shown to distribute uniformly on two concentric circles corresponding to the hydrophobic collapse and the transition to the fully folded conformations. The dependence of the distribution of the zeros on the position of the hydrophobic core, is shown to have a clear physical interpretation. The exact partition function zeros for a hairpin with a more complex structure of native contacts, the 16 C-terminal residues of streptococcal protein G B1, are also numerically computed, and their loci are also shown to be closely approximated by concentric circles. 
\end{abstract}
\pacs{87.15.ad, 87.15.hp, 87.15.Cc, 64.60.De}

\maketitle

Theoretical studies on protein folding has often been performed with simple models incorporating information on the native structure\cite{WS,M97,M98,M99,A99,G99,B00,F02,BP,D04,I07,I10}.   Wako-Saito-Mu\~noz-Eaton(WSME) model is one such example\cite{WS,M97,M98,M99}, described by Ising-like binary variables with long-range interaction on a one-dimensional lattice. The transfer matrix formalism was developed so that the exact partition function can be computed for any given temperature\cite{BP}. Since partition function contains all the information on thermodynamics, various quantities relevant for conformational transition of a protein can be calculated. However, analysis of partition function zeros\cite{YL,Fisher,IPZ,JK,AH,P03,B03,K06,Wang,CL,BDL,SYK,JL} in the complex temperature plane, one of the most powerful tools for studying phase transitions, has never been performed for the WSME model.   
The native structure has to be specified to define the Hamiltonian in  the WSME model, and we concentrate on $\beta$-hairpins in this Letter. Although a $\beta$-hairpin is a very simple structure, it captures nontrivial aspects of protein folding because contacts are formed between residues far away in sequence. Therefore, $\beta$-hairpins has been the subject of extensive researches both by experiments and computations\cite{M97,M98,M99,M06,H06,B08,J09,T09,W10,B11,W11}, including the study using WKMS model\cite{M97,M98,BP}. 
Remarkably, we find that under certain simplifying assumptions on the values of native interactions, not only the partition function, but also the partition function zeros themselves, can be obtained {\it analytically}. The exact partition function zeros for more complicated cases, which can still be obtained {\it numerically}, show similar qualitative behaviors as the analytic solution.  

 The WSME model describes a peptide or protein of length $N+1$ by an Ising-type variable $m_i\ (i=1 \cdots N)$, which denotes the state of the $i$-th peptide bond connecting $i$-th and $i+1$-th residues. The variable $m_i$ takes the value 0 or 1 depending on whether the bond is in ordered or disordered state. If the entropy of the ordered bond relative to the disordered one is denoted as $\Delta s_i < 0$, then $\lambda_i \equiv \exp(-\Delta s_i) > 0$ can be considered as the effective number of microstates of a disordered bond. 
For a $\beta$-hairpin, we assume that the local entropy cost for ordering a bond is same throughout the protein chain\cite{M97,M98,BP},  writing the effective number of disordered bond states as $\lambda = \exp(-\Delta s)$. The number of conformations of an ordered bond is 1 by definition. Note that $\lambda$ does not have to be an integer in general. 
The Hamiltonian of WSME model is
\be
H(\{m_k\}) = \sum_{i=1}^{N-1} \sum_{j=i+1}^N \epsilon_{i j} \Delta_{i j} \Pi_{k=i}^j m_k 
\ee
where $\epsilon_{i j}$ is the contact energy of $i$ and $j$-th bond, $\Delta_{i j}$ is 1 only if the bonds are in contact in the native structure and  $0$ otherwise. Thus, the contact energy is assigned if and only if the corresponding pair of bonds are in contact in the native structure, and the stretch of sequence between them are all in the ordered states. The contact energy $\epsilon_{i j}$ can either represent the backbone hydrogen bond or hydrophobic interaction between the sidechains. 
  
\begin{figure}
\includegraphics[width=\columnwidth]{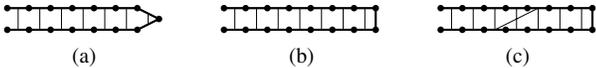}
\caption{Native conformations of simple models of $\beta$-hairpins, where the number of bonds  $N$ is  (a) even and (b) odd. The density of states can be obtained analytically, and the zeros themselves can also be obtained in analytic forms for an even value of $N$ in the limit of large $\lambda$. (c) An example of a more complex hairpin model with the lines of native contacts crossing each other. The exact density of states can be computed numerically using transfer matrix.}
\label{hairpin} 
\end{figure}

We concentrate on simplified classes of $\beta$-hairpins where the $i$-th bond form native contact only with the $N-i+1$-th bond, the one at the opposite side of the hairpin.  The native structures of these hairpins for even and odd values of $N$ are displayed in Fig.\ref{hairpin}(a) and (b), where the contacts are denoted by thin lines. The model with odd $N$ (Fig.\ref{hairpin}(b)) is more realistic in that there is an additional bond in the turn region, but the model with even $N$ (Fig.\ref{hairpin}(a)) has an advantage that an analytic formula exists for partition function zeros themselves, in the limit large $\lambda$. These models have simplifying characteristics that the lines of contacts do not cross each other.  When they do, as in Fig.\ref{hairpin}(c), the analytic formula for the density of states becomes more complicated and consequently less useful. However, one can easily compute the exact density of states and partition function zeros for a given set of parameters, using a transfer matrix\cite{BP}. We will restrict ourselves to an even value of $N$ without lines of native contacts crossing each other (Fig.\ref{hairpin} (a)), unless stated otherwise. 

 Let us call the contact between the $i$-th and $N-i+1$-th bonds as the  $i$-th contact, and  rewrite the corresponding energy as $\epsilon_{i} \equiv \epsilon_{i, N-i+1}\ (i = 1, \cdots, N/2)$ for simplicity of notation.  It is clear that the broken native contacts can appear only as a sequential stretch in the tip region, due to the restriction that the native contacts can form only when all the intervening bonds are ordered. Let us suppose that $i$-th native contacts with $i \le j$ are all broken and the rest are intact. 
The corresponding energy value is
\be
E_j = \sum_{i=j+1}^{N/2} \epsilon_i = E_N - \sum_{i=1}^{j} \epsilon_i \quad (0 \le j \le N/2)
\ee
where $E_N \equiv \sum_{i=1}^{N/2} \epsilon_i$ is the energy value of the fully folded conformation. Since at least one of the bonds forming the $j$-th contact has to be disordered,  they cannot be both in the ordered states. Therefore, the total number of states these pairs can be in is $(\lambda+1)^2-1$. All the other bonds with broken native contacts can be in any of the $\lambda+1$ states, whereas those forming the native contact is in the ordered states whose number is 1 by definition. The total number of conformations for a given value of $j$ is thus obtained by multiplying these numbers of bond states: 
\be
\Omega(E_j; \lambda) = \left\{
    \begin{array}{ll}
        1 &  (j=0),\\
        \left((\lambda+1)^2-1\right)(\lambda+1)^{2j-2} &  (1 \le j \le N/2)
    \end{array} \right. \label{DS}
\ee  
where $j=0$ corresponds to the fully folded conformation. 
If all the native contacts are due to hydrogen bonds, we may assign equal energy value $\epsilon_i = \epsilon < 0$ to each contact, and the partition function for an even value of $N$ is obtained in analytic form from Eq.(\ref{DS}) as a function of $z \equiv e^{\beta \epsilon}$:
\be
Z\! &=&\! z^{-N/2} \frac{\lambda^2 + 2 \lambda}{(\lambda+1)^2} \left [ \frac{(\lambda + 1)^2}{\lambda^2 + 2 \lambda} + \sum_{j=1}^{N/2} \left((\lambda + 1)^2 z \right)^j \right] \label{anapart}
\ee 
When $\lambda$ is large enough so that
\be
\frac{\lambda^2 + 2\lambda + 1}{\lambda^2 + 2 \lambda} \simeq 1, \label{approx}
\ee 
we may approximate the partition function as
\be
Z \simeq z^{-N/2} \frac{\lambda^2 + 2 \lambda}{(\lambda+1)^2} \left [ \sum_{j=0}^{N/2} \left((\lambda + 1)^2 z \right)^j \right], \label{apppart}
\ee
so that the solution to the equation $Z(z)=0$ is obtained analytically as
\be
z_j = \frac{1}{(\lambda+1)^2} \exp \left(\frac{2 \pi i j}{N/2+1}\right) \quad (j=1 \cdots N/2). \label{anazero}
\ee
That is, the zeros are uniformly distributed along the circle of radius $1/(\lambda+1)^2$ except for the point on the positive real axis. Since the physical region is $0<z<1$, we see the folding transition exists as long as $\lambda > 0$, according to Eq.(\ref{anazero}), with corresponding folding temperature at $T_f = - \epsilon/(2 k_B \ln (\lambda+1))$. The folding temperature decreases  as $\lambda$ increases, since the unfolded conformation becomes more favored entropically. As is well known, the uniform distribution of zeros on a circle leads to a first-order transition in the limit of infinite $N$\cite{YL}.

\begin{figure}
\includegraphics[width=\columnwidth]{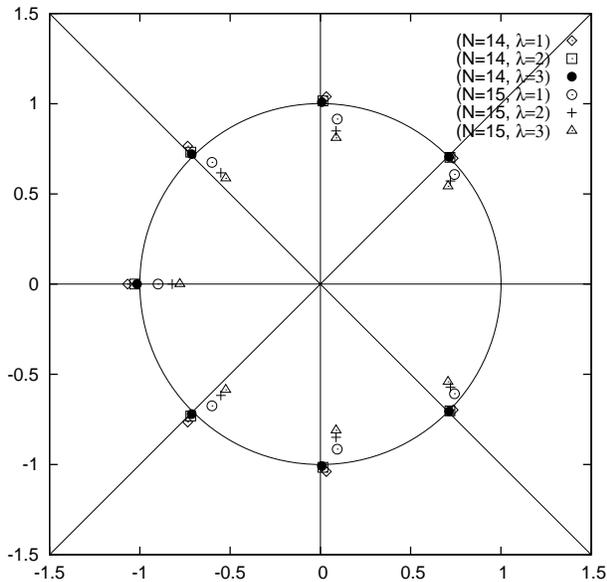}
\caption{The partition function zeros of WSME model of $\beta$ hairpin models with 7 native contacts  
and uniform interaction strengths, in the complex plane of $(\lambda+1)^2 z$.  The analytic solution Eq.(\ref{anazero}) lies on the intersection of the circle and the straight lines, which fits quite well with the numerical solutions for even chain length $N=14$. The zeros for odd chain length $N=15$ are also plotted.}
\label{circle1} 
\end{figure}

Although the approximation Eq.(\ref{approx}) is better for larger $\lambda$, the locations of zeros fit quite well with the analytic solution Eq.(\ref{anazero}) also for small values of $\lambda$, even for the extreme case $\lambda = 1\ (\Delta s=0)$, as can be seen in Fig.\ref{circle1} \footnote{Note that even for vanishing local entropy $\Delta s=0$ for disordered bond, the unfolded conformation has higher entropy since both ordered and disordered bonds contribute to the ensemble of unfolded conformations.}. In the figure, the partition function zeros for $N=14$, obtained by solving the polynomical equation Z(z)=0 by MATHEMATICA, are plotted on the complex plane of $(\lambda+1)^2 z$ for several integer values of $\lambda$.  
The straight lines at angular interval of $\pi/4$ intersecting are also drawn along with the unic circle, so that their intersection are are the analytic solutions Eq.(\ref{anazero}), except for the one on the positive real axis. We see the distances of the zeros from the origin are slightly larger than the radius of the circle, because the entropic cost of the fully folded state is overestimated in  the approximation Eq.(\ref{approx}). Note however that the deviation is visible only near the negative real axis, and almost negligible near the positive real axis which is the physically meaningful region.
 
For odd values of $N$, the density of states is almost the same as that of the even $N$ with the same number of native contacts, except that the number of the fully unfolded states is larger,
\be
\Omega(E_{(N-1)/2}) = \left( (\lambda+1)^3 - 1 \right)(\lambda+1)^{N-3}, 
\ee
due to the contribution of the extra bond at the turn which is liberated. The zeros for $N=15$, corresponding to the same number of native contacts as $N=14$, are also plotted in Fig.(\ref{circle1}). As can be seen from the figure, they are located inside the circle, due to the fact that now the fully unfolded structure is more favorable compared to that of even $N$, because of additional entropic contribution from the extra bond at the turn. The deviation from the even length chains increases as $\lambda$ increases, as to be expected.

\begin{figure}
\includegraphics[width=\columnwidth]{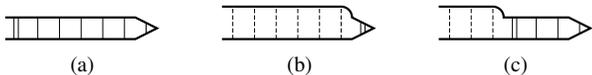}
\caption{Intermediate conformation where a hydrophobic core (double line) is formed but the tip region is unfolded. The dashed lines denote broken native contacts. (a) When a hydrophobic core exists near the tip of the hairpin, it can be formed only in the fully folded conformation. (b) When the hydrophobic core is near the turn, the intermediate conformation is almost indistinguishable from unfolded conformation in terms of energy and entropy. (c) Nontrivial folding intermediates can be formed when the hydrophobic core is located near the middle of the hairpin.}
\label{intermed} 
\end{figure}

\begin{figure}
\includegraphics[width=\columnwidth]{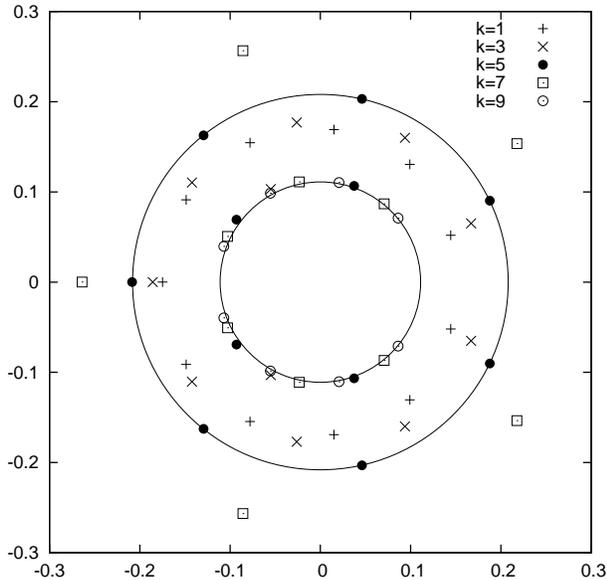}
\caption{Partition function zeros of WSME model of $\beta$ hairpin models with a hydrophobic core. There are three additional zeros for $k=9$ outside the range of the plot, at $z=-0.45$ and $z=0.28 \pm 0.41 i$.}
\label{circle2} 
\end{figure}

By introducing hydrophobic interaction in addition to the hydrogen bond, we can observe the collapse transition to an intermediate where the hydrophobic core is formed but the tip region is unfolded(Fig.\ref{intermed}(c)). 
Again, the partition function zeros can be obtained analytically for a special case. Consider the case where $N$ is even, $n \equiv N/2$ is odd and the hydrophobic core consists of a single additional interaction at $h \equiv (n+1)/2$, with (free) energy $\Delta G_{SC} = q \epsilon< 0$ with some integer $q$. Also, we assume that each hydrogen bond contributes the energy $\Delta H_{hb} = p \epsilon <0$ where $p$ is also an integer.
The partition function can then be factorized as
\be
 Z &=& z^{-pn-q} \frac{\lambda^2 + 2 \lambda}{(\lambda+1)^2} [ \frac{\lambda^2+ 2\lambda + 1}{\lambda^2 + 2 \lambda} + \sum_{j=1}^{h-1} (\lambda + 1)^{2j} z^{p j}\nonumber\\ 
&&+  \sum_{j=h}^n (\lambda + 1)^{2j} z^{p j + q} ]\nonumber\\
&\simeq& z^{-pn-q} \left [ \sum_{j=0}^{h-1} \left((\lambda + 1)^2 z^p \right)^j \right] \nonumber\\
&&\times\left[ 1 + (\lambda+1)^{2 h} z^{h p + q} \right]
\ee
where the approximation (\ref{approx}) is used as before. The first and the second factors give two concentric circles for zeros:
\be
z_j &=& \frac{1}{(\lambda+1)^{2/p}} \exp (\frac{2 \pi i j}{h p}) \quad (j=1, \cdots, hp -1)\nonumber\\
\tilde z_j &=& \frac{\exp (\frac{(2j+1) \pi i}{h p + q})}{(\lambda+1)^{2h/(hp + q)}}  \ (j=0, \cdots, hp+q-1),
 \label{conzero}
\ee 
again showing the first-order-like nature of the transition.
It is easy to see from the analytic solution of the loci (\ref{conzero}), that the folding and collapse transition occurs at $T_f = - \epsilon p/[2 k_B \ln (\lambda+1)]$ and $T_c = -\epsilon (hp + q)/[2 k_B h \ln (\lambda+1)]$.
As to be expected, larger value of $\Delta H_{hb}$ corresponds to higher value of $T_c$ and denser distribution of zeros on the outer locus signifying sharper collapse transition. Also, for $q=0$ the two circles collapse to one circle corresponding to the folding transition, Eq.(\ref{anazero}), by setting $p=1$ without loss of generality. The exact partition function zeros for $n=9$, $p=1$, $q=2$, and $\lambda=2$, with an extra hydrophobic interaction at the $k$-th contact, are plotted on complex $z$-plane, in Fig.\ref{circle2}, along with the
circles at radii $1/(\lambda+1)^{2/p} = 1/9 \simeq 0.111 $ and $1/(\lambda+1)^{2h/(hp+q)} = 1/9^{5/7} \simeq 0.208$.  
We see that the zeros for $k=5$ are extremely well described by the analytic solution (\ref{conzero}), being distributed on inner and outer circles at angular interval of $2\pi/hp = 2\pi/5$ and $2 \pi/(hp+q) = 2 \pi/7$ respectively.  We see that as the position of the hydrophobic core is moved toward the tip, the radius of the outer locus decreases since the intermediate becomes unfavorable entropically. Also, the density of zeros of the inner locus decreases, due to the fact that the intermediate and the fully folded conformation becomes less distinguishable(Fig.\ref{intermed}(a)). Eventually, at $k=1$ the zeros form one locus corresponding to the folding transition. On the other hand, as the hydrophobic core moves toward the turn, the radius of the outer locus increases and its density decreases, because the entropy of the intermediate increases and it becomes less distinguishable from the unfolded state(Fig.\ref{intermed}(b)). 

\begin{figure}
\includegraphics[width=\columnwidth]{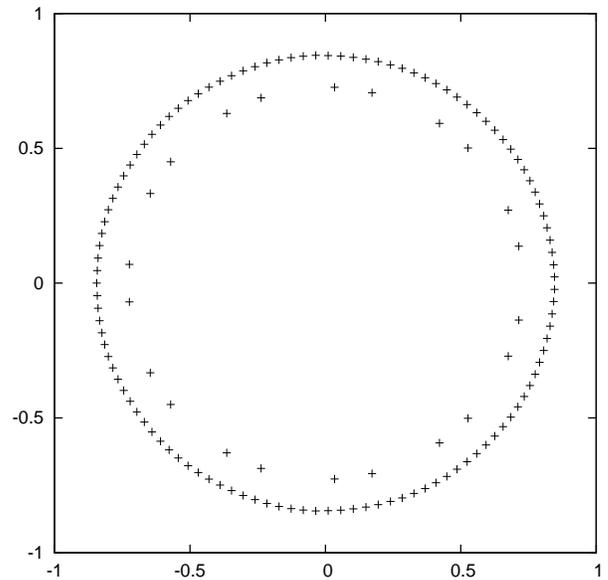}
\caption{The partition function zeros of WSME model of $\beta$ hairpin, 16 C-terminal residues of streptococcal protein G B1.}
\label{beta} 
\end{figure}

So far we have concentrated on general class of simplified hairpin models. Real $\beta$-hairpin, 16 C-terminal residues of streptococcal protein G B1 (GB1), was also studied with WSME model\cite{BP}, which includes crossed lines of native contacts (Fig.\ref{hairpin}(c)). The density of states was calculated using the transfer matrix formalism\cite{BP}, where the native contacts are given in ref.\cite{M98}. The hydrogen bond and hydrophobic interaction energies are $\Delta H_{hb} = -1.1\ {\rm kcal/mol}$, $\Delta G_{SC} = -2.0\ {\rm kcal/mol}$, and the local entropic cost of folding is $\Delta s = -3.12\ {\rm cal/K mol}$, which corresponds to $p=11$, $q=20$ with $\epsilon=-0.1\ {\rm kcal/mol}$, and $\lambda=4.80$. The zeros are obtained as solutions of a 137-th order polynomial equation, which are plotted in the plane of $e^{\beta \epsilon}$  in Fig.(\ref{beta}). We see that the loci of zeros also form concentric circles with radii 0.727 and 0.845 corresponding to temperatures $T=158$ K and $T=299$ K. Note however, that the angular distribution of inner circle is not uniform. The transition temperature $T=299$ K is quite consistent with 297 K defined in terms of kinetic rates and free energy in refs.\cite{M97,BP}. A two state transition behavior at this temperature was reported experimentally\cite{M97}, and also confirmed by theoretical study using WSME model\cite{BP}. In fact, since the distribution of the zeros on the inner locus is very sparse, especially near the positive real axis, any possible transition from hydrophobically collapsed intermediate to the fully folded state near $T \sim 158$ K might be quite smooth, making it hard to be observed as a meaningful conformational transition.  

Partition function zeros of protein models with complex structure of inter-residue contacts can also be studied using the transfer matrix formalism, which is left for future study.

The author thanks Seung-Yeon Kim and Jae Hwan Lee for useful discussions.

\end{document}